# Influence by trajectorial electron transport on anomalous ultrasound attenuation in high pure Gallium single crystal

A. G. Shepelev, O. P. Ledenyov and G. D. Filimonov

*National Scientific Centre Kharkov Institute of Physics and Technology,
Academicheskaya 1, Kharkov 61108, Ukraine.*

The anomalous propagation of longitudinal ultrasound with the frequencies of *70* and *150 MHz* in the high pure *Gallium* single crystal at the weak external magnetic field ($H \perp k$) at the low temperature of *0.4 K* was researched experimentally. The delayed ultrasonic signal, comparing to the main ultrasonic signal, is detected. The research on the new magneto-acoustic effects made it possible to find the temperature dependence of frequency of collisions between the electrons and the thermal phonons, $v_{e,ph}(T) \sim T^3$, in the high pure *Gallium* single crystal. It is found that the anomalous oscillations of propagation velocity of ultrasonic signal pulses and the signals, propagating by the way of the electron transport in the high pure *type I* superconductor, can be detected at the directions of the magnetic field $H$ close to $H \parallel k$.

## Introduction

The first magneto-acoustic researches on the electron properties of an intermediate state in the high pure *Gallium* single crystal were published around *15* years ago [1]. There are more than the *30* research publications on the subject of research interest by the *American*, *French*, *Russian*, *Ukrainian* and *Canadian* researchers (the early published research works are reviewed in [2]). In the recent time, the special interest in the research toward the nature of anomalous ultrasound propagation in an intermediate state of the high pure *type I* superconductors at the weak magnetic fields has appeared in [3-5].

The authors of present research discovered the anomalously big oscillations of the velocity of the ultrasonic pulses propagation in the high pure *Gallium* single crystal at the weak magnetic fields at the temperature $T \simeq 0.4 K$ in [3]. The strong periodic changes of the time position of the propagated-through-the-sample ultrasonic signal impulse at the change of the orientation or magnitude of the magnetic field $H$, $H > H_C = 50\ Oe$, in the high pure *Gallium* single crystal were observed and considered as the new effect of the anomalously big oscillations of the ultrasound velocity at the ultrasound propagation in an intermediate state of the high pure *type I* superconductor in the weak magnetic fields at the low temperatures.

Our further researches showed that, at the change of the orientation or magnitude of the magnetic field $H$, the shape of the ultrasonic signal impulse changes to some degree in addition to the observed strong periodic changes of the time position of the propagated-through-the-sample ultrasonic signal impulse. In the subsequent experiments, conducted by *Fil', Burma, Bezuglyi* [4] at the similar experimental conditions, the precursor-signals, which propagated through the sample much faster than the main ultrasonic signal impulse, were observed in an intermediate state of the high pure *type I* superconductor at the smaller magnitudes of the magnetic field $H$. The theoretical explanation by *Bogachek*, *Rozhavsky*, *Shekhter* [5] is based on the theoretical proposal about the possible transfer of the ultrasonic signals by the electrons on the distances, which are multiple to the diameter of electron orbit in the magnetic field $H$ – the splashes of the sound field in the superconductor, which are analogous to the anomalous penetration of the ultra high frequency electromagnetic field [6]. In this case, the discovered effect [3] must be interpreted as a result of the redistribution of the energy by the electrons inside the ultrasonic signal impulses package, when the diameter of the electron orbit is smaller than the spatial length of the ultrasonic signal impulse, hence the main ultrasonic signal impulse and newly originated ultrasonic signal impulses (at the ultrasonic signal transfer by the electrons) are not divided. The superposition of these three signals (the precursor signal, delayed signal and main signal) defines the nature of the anomalous oscillations of propagation velocity of ultrasonic signal impulses.

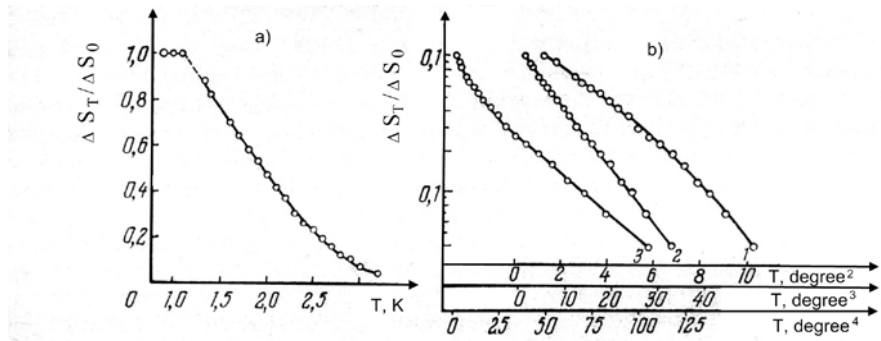
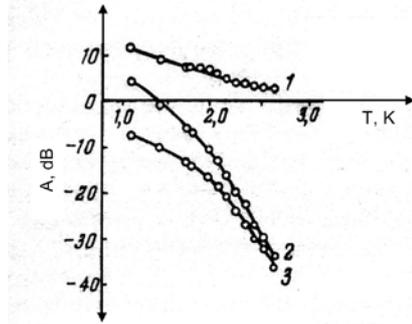
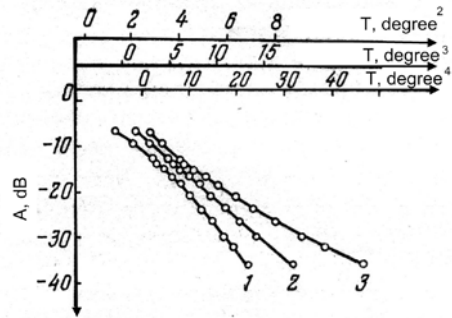

Fig. 1 Highest amplitude of oscillations of ultrasound propagation velocity $\Delta S_T/\Delta S_0$ in high pure Ga single crystal at longitudinal ultrasonic signal frequency of 60 MHz in plane (**a**, **c**) at external magnetic field H=48.6 Oe at **k** ∥ **b**, **H** ⊥ **k** as function of:
a) Temperature T;
b) Temperature $T^2$ (curve 1), Temperature $T^3$ (curve 2), Temperature $T^4$ (curve 3).

Fig. 2 Temperature dependence of amplitudes of:
Main ultrasonic signal pulse (curve 1), Precursor ultrasonic signal pulse (curve 2), Precursor ultrasonic signal pulse with consideration of temperature change of main ultrasonic signal pulse (curve 3) in high pure Ga single crystal at longitudinal ultrasonic signal frequency of 150 MHz at H = 30 Oe, ∡H, a=44°, **k** ∥ **b**, **H** ⊥ **k**.

Fig. 3 Dependence of amplitude of precursor ultrasonic signal pulse (at logarithmic scale) on Temperature $T^2$ (curve 1), Temperature $T^3$ (curve 2), Temperature $T^4$ (curve 3) in high pure Ga single crystal.

**Discussion on measurements results**

In this research, the more detailed experimental results on the possible influence by the trajectorial electron transport on the anomalous ultrasound attenuation in an intermediate state of high pure *Gallium* single crystal in the weak magnetic field at the low temperatures are presented. The experimental measurements setup and the high precision measurements methodology for the accurate characterization of the high pure *type I* superconductors in the weak magnetic fields at the low temperatures, using the impulse method, are described in [3].

1. The massive sample of high pure *Ga* single crystal with the length of *11 mm*, having the orientation of **k** ∥ **b** – axis of the *Ga*, **H** ⊥ **k**, is researched with the purpose of the increase of amplitude of ultrasonic impulse. It is found that the magnitude of effect of anomalously big oscillations of velocity of ultrasonic impulses propagation does not depend on the amplitude of ultrasonic signal. Besides, there are the two weak ultrasonic impulses, propagating with the faster velocity (the precursor signal) and the slower velocity (the delayed signal) in comparison with the main ultrasonic impulse propagation velocity, which appear in the high pure *Ga* single crystal at the lower magnitudes of the external magnetic fields *H* than the magnitude of the critical magnetic field $H_c$ at the temperature $T > Tc = 1.08 K$. The magnitude of delayed impulse is in a few times smaller than the magnitude of precursor signal, because the delayed signal propagates on the additional distance in the superconductor. The detection of delayed impulse is in agreement with the theory [5].

2. It was in the scope of our research interest to determine the influence by the temperature change on the anomalous propagation of ultrasonic impulse in an intermediate state of the high pure *Ga* single crystal at the weak magnetic fields at the low temperatures. To achieve our research goals, the oscillatory dependence of the ultrasonic signal propagation velocity on the orientation of the magnetic field ***H*** in relation to the axis ***a*** (in the plane: ***H*** ⊥ ***k***) was researched in a sample of the high pure *Ga* single crystal with the length of *20 mm* at the ultrasonic signal frequency of *70 MHz* at the ultrasonic signal impulses length of *1 μsec* at the constant magnitude of the external magnetic field *H = 48.6 Oe*.

In Fig. 1, the temperature dependence of maximal amplitude of oscillations of resulting ultrasonic signal propagation velocity $\Delta S_T / \Delta S_0$ is shown in the auxiliary units. The magnitude of effect does not depend on the temperature *T* below the temperature of *1.1 K*, because the electron mean free path *l* does not change in the high pure *Ga* single crystal. It can be seen that the part of the effect, which depends on the temperature *T*, is described by the expression

$$\Delta S_T / \Delta S_0 \sim exp(-\beta \cdot T^3).$$

The experimental measurements data on the temperature dependence of the precursor signal in a sample of the high pure *Ga* single crystal with the length of *11 mm* were also obtained. The measurements were conducted at the ultrasonic signal frequency of *150 MHz* with the ultrasonic signal impulse length of *0.5 μsec* at the constant magnitude of magnetic field *H = 30 Oe* with the orientation of magnetic field ∡*H, a = 44°*.

In Fig. 2, the measured temperature dependences of amplitudes of main ultrasonic signal impulse, propagated through a sample (curve 1), precursor ultrasonic signal impulse (curve 2), and precursor ultrasonic signal impulse with the consideration of temperature change of main ultrasonic signal impulse (curve 3) are shown. During the experimental measurements, the relations between the above mentioned amplitudes and the amplitude of the ultrasonic signal impulse at the magnetic field *H=0*, which does not depend on the temperature, because of the big magnitude of the parameter *k·l* at the wide range of temperatures below *4.2 K*, were found.

In Fig. 3, the dependence of the amplitude of precursor ultrasonic signal impulse on the temperature $T^3$ is presented; it can be seen that the amplitude of precursor ultrasonic signal impulse is

$$A(T) \sim exp(-\beta \cdot T^3).$$

Let us suppose that the amplitude of the precursor ultrasonic signal impulse as well as the amplitude of the radio-frequency dimensional effect [7, 8] are equal to

$$A \sim exp(-\lambda/l) = exp(-\pi \cdot v/\Omega),$$

then the temperature dependence of the precursor ultrasonic signal impulse is

$$A(T) \sim exp[-\pi \cdot v(T)/\Omega],$$

where *λ* is the path by the electron along its trajectory until the scattering, *l* is the electron mean free path, *v* is the frequency of collisions, and *Ω* is the electron cyclic frequency.

The dependence $v(T) \sim T^3$, which was detected at the research of the new effects, is probably defined by the electron scattering on the phonons in the crystal grating $v_{e, ph}$*). It is assumed that this dependence can be understood by taking the following things to the consideration. Despite of the fact that the ultrasonic waves package with the length *L*, which is close to the electron orbit diameter *D*, is transported by the electrons; the value $1/2\lambda_{sound}$ plays the role of the characteristic distance, where the phase of ultrasonic excitation is changed; that is the value, which is in many times smaller than the *L* and *D* ($\lambda_{sound}$ is the wavelength of ultrasonic signal). This circumstance results in the situation, when the electrons can not effectively interact with the ultrasound, because of the presence of small angles scattering by the electrons on the thermal phonons [9] as in the case of the usual magneto-acoustic effects [8, 10].

It is interesting to note that the small change of orientation of magnetic field in the plane ***H*** ⊥ ***k*** results in the significant change of the amplitude of the precursor ultrasonic signal impulse in comparison with the amplitude of the main ultrasonic signal impulse. In agreement with the theory [5], it can be considered as an evidence of the anisotropy of deformation potential. The expression of the temperature dependence of the collisions frequency is the same as it was before $v(T) \sim T^3$, the coefficient *β* changes only (the tilt of curve 2 in Fig. 3).

3. Let us comment on the two discovered features:

a) There is the dimensional effect of electron orbits cutoff – the precursor ultrasonic signal appears at the lower magnetic fields in the bigger samples - in the various samples of the high pure *Ga* single crystal with the same orientation of crystallographic axes.

b) In distinction from the theory [5], which was created for the parallel orientation of the magnetic field ***H*** in relation to the metallic plate and the perpendicular orientation of the magnetic field ***H*** in relation to the wave vector ***k***; there are the precursor ultrasonic signal and the oscillations of the propagation velocity of the ultrasonic signal impulses at the big tilting angles of the magnetic field ***H*** in relation to the plane ***H*** ⊥ ***k***. Moreover, the effects can be observed at the direction of the magnetic field ***H*** close to ***H*** // ***k***. Probably, the nature of the discovered effects is more complex than the first proposed theoretical description [5]. It is possible to assume that the anomalous propagation of the ultrasonic signal in an intermediate state of the high pure *Ga* single crystal is a wide spread phenomenon, which can be observed in the other high pure *type I* superconductors at the specified conditions.

## Conclusion

The experimental research on the anomalous propagation of longitudinal ultrasonic signal at the frequencies of *70* and *150 MHz* in the high pure *Gallium* single crystal at the weak external magnetic field (***H*** ⊥ ***k***) at the low temperature of *0.4 K* is completed.

The delayed ultrasonic signal, comparing to the main ultrasonic signal, is detected. The research on the new magneto-acoustic effects helped to find the temperature dependence of frequency of collisions between the electrons and the thermal phonons, $v_{e,ph}(T) \sim T^3$, in the high pure *Gallium* single crystal. It is found that the anomalous oscillations of propagation velocity of ultrasonic signal pulses and the signals, propagating due to the electron transport in the high pure *type I* superconductor, can be detected at the directions of the magnetic field ***H*** close to ***H*** // ***k***.

Authors are very grateful to Boris G. Lazarev, E. A. Kaner, V. F. Gantmakher and A. M. Kosevich for the interesting thoughtful discussions on the experimental research results.

This research paper was published in the *Journal of Low Temperature Physics* (*FNT*) in 1976 in [11].

*) The simple dependence $v(T) \sim T^3$, despite the existence of the big electron mean free path in the high pure *Ga* single crystal, resulting in the possibility of multiple interactions between the electrons and the ultrasound oscillations before the electron scattering, is probably stipulated by the change of the phase of ultrasound oscillations at the time of electron rotation along the orbit between the interactions.

1) The experimental research results were reported at the *IX* conference on the acoustic electronics and quantum acoustics, Moscow, Russia, June 24, 1976.

*E-mail: ledenyov@kipt.kharkov.ua